# Cryogenic Memory Technologies

Shamiul Alam[1], Md Shafayat Hossain[2], Srivatsa Rangachar Srinivasa[3], and Ahmedullah Aziz[1*]

[1]Dept. of Electrical Eng. & Computer Sci., University of Tennessee, Knoxville, TN, 37996, USA
[2]Dept. of Physics, Princeton University, Princeton, NJ, 08544, USA
[3]Intel Labs, Hillsboro, OR, 97124, USA
*Corresponding Author's Email: aziz@utk.edu

*Abstract*—The surging interest in quantum computing, space electronics, and superconducting circuits has led to new developments in cryogenic data storage technology. Quantum computers promise to far extend our processing capabilities and may allow solving currently intractable computational challenges. Even with the advent of the quantum computing era, ultra-fast and energy-efficient classical computing systems are still in high demand. One of the classical platforms that can achieve this dream combination is superconducting single flux quantum (SFQ) electronics. A major roadblock towards implementing scalable quantum computers and practical SFQ circuits is the lack of suitable and compatible cryogenic memory that can operate at 4 Kelvin (or lower) temperature. Cryogenic memory is also critically important in space-based applications. A multitude of device technologies have already been explored to find suitable candidates for cryogenic data storage. Here, we review the existing and emerging variants of cryogenic memory technologies. To ensure an organized discussion, we categorize the family of cryogenic memory platforms into three types – superconducting, non-superconducting, and hybrid. We scrutinize the challenges associated with these technologies and discuss their future prospects.

## 1. Introduction

Silicon-based complementary metal-oxide-semiconductor (CMOS) technology now allows us to accommodate more than billion transistors on a chip (for example, Cerebras Wafer Scale Engine 2 with 2.6 Trillion transistors [1]). However, this aggressive scaling comes with a price- an exorbitant power dissipation reaching the physical limit. For example, in 2018, about 205 TWh of electricity was consumed by the US data centers [2]. To circumvent this issue, numerous 'beyond-CMOS' technologies, such as superconducting single flux quantum (SFQ) electronics, are being explored. SFQ circuits and systems based on superconductive devices [e.g., Josephson junctions (JJ) and superconducting quantum interference device (SQUID)] are generally faster and more energy-efficient than the CMOS counterparts[3], thanks to the dissipation-less current flow in superconductors. Ultrafast and energy-efficient superconducting SFQ circuits, coupled with lossless and low dispersion interconnects, have significantly improved performance in different applications such as digital radio frequency receivers [4,5], high-end computing [6], and so on. However, the lack of fast, low-power, high-density cryogenic memory compatible with the performance of SFQ circuits is one of the significant challenges for implementing practical and reliable SFQ systems.

Going beyond the conventional computing paradigm, the importance of cryogenic memory is felt in quantum computers as well. Among the different platforms for quantum computations, superconducting qubits have attracted immense attention thanks to their non-dissipative and strongly non-linear behavior [7–16]. Quantum computers promise to solve commercially and scientifically important problems that the current classical computers can hardly handle in a realistic timeframe. For instance, accurate simulation of large molecules for the development of new drugs and eco-friendly manufacturing processes now require unrealistic time due to exponential scaling of the complexity in classical computers. However, these simulations can be performed with a quantum computer in polynomial time [17]. Moreover, quantum computers can significantly accelerate number theory, algebraic, and optimization problems [18]. However, the lack of a suitable cryogenic memory is a major challenge in implementing a scalable quantum computer.

Space-based application is another major field that can highly benefit from the utilization of cryogenic devices. JJ and SQUID-based cryogenic equipment have been used in several spacecrafts in the last few decades [19]. The utilization of cryogenic devices has enabled the spacecrafts to work with the electromagnetic radiation emitted by celestial objects over a wavelength range that is difficult to work with from the ground. The utilization of cryogenic electronics is expected to improve efficiency and reliability and simplify the design of the space systems, thanks to the improvement in electrical, electronic, and thermal properties of materials at cryogenic temperatures [20,21].

In this review, we will discuss state-of-the-art cryogenic memory technologies. We organize the review as the following: *Section 2* delineates the benefits of using cryogenic memory in quantum computers and cryogenic electronics, as well as the required characteristics of a cryogenic memory for these applications. Next, in *Section 3*, we describe different cryogenic memories, namely cryogenic non-superconducting memories [e.g., charge-based memories and resistance-based memories], superconducting memories [such as superconductor-insulator-superconductor (SIS) JJ-based, magnetic JJ-based, superconducting memristor-based, and ferroelectric SQUID-based approaches], and hybrid memories (utilizes both superconducting and non-superconducting technologies). In *Section 4*, we present a comparative study on the progress made so far on the cryogenic memory technologies. Here we also outline the major challenges faced by different cryogenic memories and conclude with the research prospects on developing a suitable cryogenic memory.

## 2. Why Cryogenic Memory?

Cryogenic memories are critical for quantum technologies. A typical quantum computer has three major components: quantum substrate (qubits), a control processor, and a memory block [8] [see Fig. 1(a) which illustrates the organization of a scalable quantum computer proposed in Ref. [22,23]]. In quantum computing, qubit is the basic unit of quantum information realized with a two-state device. To protect the quantum states of the noise sensitive qubits, they are placed at a few milli-Kelvin temperature[10–15]. In the current lab size quantum computers, a conventional computer at room temperature (300 K) is used as the control processor. Long control cables are required to connect the qubits and the control processor. Although this architecture works fine for a small number of qubits, it cannot be scaled to a few hundreds of qubits because it would require a large number of wires to establish a connection between the room-temperature control processor and the qubits at cryogenic temperature [8]. Note, the quantum computer needs to be scaled up to thousands of qubits to utilize its full potential. For example, to run Shor's algorithm for prime factorization[18], which is a textbook quantum algorithm, on a 1024-bit number in a quantum computer, two quantum registers (one with 2048 qubits and another with 1024 qubits) are required. To circumvent this scaling issue, the control processor should be placed at a temperature very close to that of the qubits using the superconducting interconnects[24]. Also, the use of superconducting interconnects instead of metal wires will also be beneficial because of the dissipationless nature and ability of high density of the superconducting interconnects. However, to use superconducting interconnects, the control processor needs to be placed at 4 K or below. It places yet another restriction in the type of memory that can be used in conjunction with quantum computers: room temperature memories cannot be used because the interconnects between the room-temperature memory and the cryogenic control processor would have a significant thermal leakage. This leakage stems from the huge temperature difference (about 300 K) between the room and cryogenic temperature and contributes to thermal noise that is large enough to destroy the quantum states of the noise-sensitive qubits [8]. An obvious way-around is to use a cryogenic 4 K memory, which is compatible with the superconducting control processor.

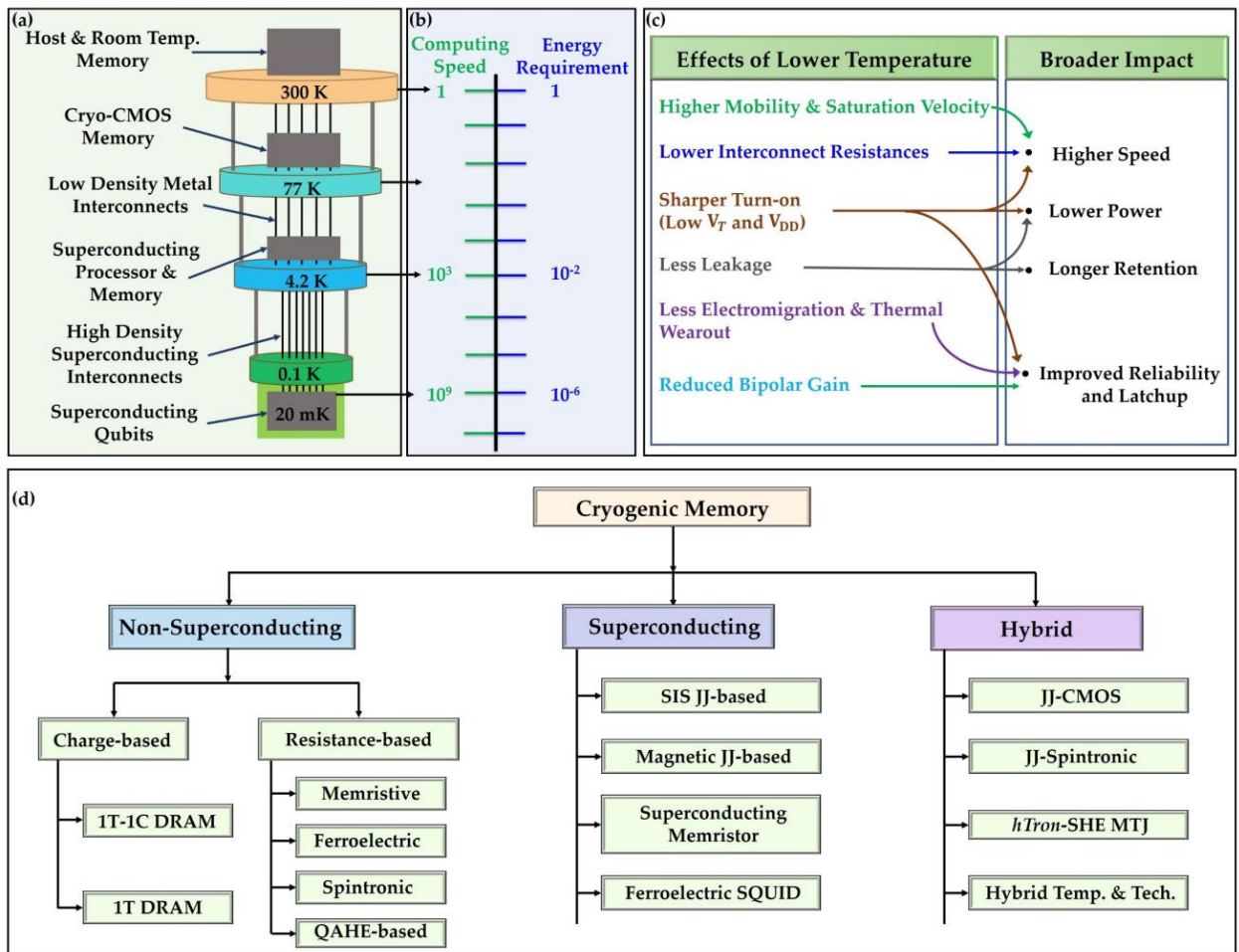

**Fig. 1: (a)** Proposed organization of a superconducting quantum computer [22,23]. Control processor operating at 4 K provides instructions to the qubits operating at 20 mK. Possible memory locations are 4 K, 77 K and 300 K. **(b)** Significant increase in speed and decrease in energy with the lowering of operating temperature. Speed and energy have been normalized by the values at 300 K. **(c)** Effects of cryogenic temperature on the traditional CMOS devices. Lower temperature improves the speed, power requirement, retention time and reliability of the CMOS memories [29]. **(d)** Taxonomy of state-of-the-art cryogenic memory technologies.

Another key requirement for the storage device of a quantum computer is its high storage capacity. The state-of-the-art quantum algorithms require a large number of arbitrary rotations which eventually necessitate a large program memory [8]. Furthermore, to preserve the data integrity of the qubit states, the qubits undergo continuous error correction schemes [8,11]; it requires extensive memory and bandwidth.

Although 4 K memories promise better scalability of the quantum computer and very low thermal leakage, researchers have explored the option of placing memories at a temperature higher than the processor temperature because the cost for cooling the memory systems down to 4 K is $\simeq$ 200 times higher compared to the room temperature memory [10,22]. Albeit there is a small thermal leakage, memories placed at a higher temperature (such as 77 K) than 4 K requires lower cost [10]. Figure 1(a) highlights all these memory placements in a quantum computer with superconducting qubits.

Cryogenic memories are also crucial for superconducting SFQ electronics. To solve the speed and power issues faced by CMOS technology [25], superconducting electronics is an attractive solution, thanks to the faster and more energy-efficient operation offered by superconducting devices (JJs and SQUIDs). Figure 1(b) depicts the significant reduction in power requirement and increase in speed with the lowering of operating temperature. However, according to the National Security Agency, the lack of a scalable and compatible 4 K memory is one of the major challenges that has limited the superconducting electronics only in niche applications [6]. The reason for the requirement of a 4 K memory in SFQ technologies is the lower power and latency afforded by the physical proximity [6].

In the next section, we list the frontrunners in the cryogenic memory technologies and discuss how they perform compared to the requirements mentioned above in this section.

### 3. Overview of State-of-the-art Cryogenic Memories

Here, we highlight the state-of-the-art cryogenic memory architectures- charge and resistance-based non-superconducting memories, JJ and SQUID-based superconducting memories, and hybrid memories implemented using both non-superconducting and superconducting technologies. For ease of discussion, we first categorize the state-of-the-art cryogenic memories as shown in Fig. 1(d).

### 3.A. Cryogenic Non-Superconducting Memories

Our discussion starts with the non-superconducting memories for cryogenic applications. These memory devices are the prime candidates for cryogenic memory owing to their technological maturity and their ability to offer high storage capacity. Remarkably, there is a recent report of a CMOS control processor that can operate at 4 K[11]. It permits a close integration between superconducting qubits and CMOS control processors. Then the remaining question is whether or not the conventional non-superconducting memories work in cryogenic temperatures.

Yoshikawa *et al.* [26] suggested that CMOS devices and circuits manufactured using several sub-micron CMOS processes operate better at 4 K than the room temperature. The speed improves by 40% to 50% and the power dissipation reduces by 30% (depending on the circuit operation) for digital circuits at 4 K compared to the room temperature [26]. However, operating at 4 K temperature poses entails significantly higher costs [10,22]. Therefore, operating at a slightly higher temperature can offer an optimized solution to this challenge.

As shown in Fig. 1(a), 20 mK, 4 K, 77 K, and 120 K are the temperature levels of a typical cryogenic (dilution) refrigerator [8]. A natural choice is to run the CMOS memory operations at 77/120 K. These temperatures still reduce the leakage current substantially and improve the carrier mobility [27], leading to an enhancement in the driving capability of access transistors and the overall memory operation [Figure 1(c) shows the improvement in different performance metrics of CMOS devices with the lowering of temperature]. Importantly, these temperatures are within the ideal operating temperature range for the CMOS devices. In turn, placing a memory at a slightly higher temperature than 4 K cuts down the cost substantially. While this scheme may work for smaller quantum computers, it will most certainly limit the scaling of the quantum computer due to the large number of connections between the 77/120 K memory and the 4 K control processor (and the resulting thermal loss). Moreover, some emerging memories have been reported to show successful operation at 4 K temperature. So, for the non-superconducting memories, the main challenge is their higher power demand and lower speed compared to the control processor at 4 K.

To facilitate our discussion, we classify the non-superconducting memories into two major categories: (i) charge-based and (ii) resistance-based. In the following subsections, we will introduce different members of these two families.

### 3.A.1. Charge-based Memories

First, we discuss the different variants of CMOS dynamic random-access memory (DRAM). In the charge-based memories, data is stored in a small capacitor. The capacitor can either be charged or discharged which provides the two memory states ('0' and '1') in the cell. There are two main categories of DRAMs: DRAM with capacitor [mostly known as 1-Transistor 1-Capacitor (1T1C) DRAMs] and capacitorless DRAM [known as 1-Transistor (1T) DRAMs].

### 3.A.1.1. 1-Transistor 1-Capacitor (1T1C) DRAMs

1T1C DRAMs [see Fig. 2 (a)] are one of the strong candidates for cryogenic memory because of their technical maturity, the ability to offer high storage capacity, and expected performance improvements at low temperatures due to the reduced junction leakage of cell transistors [8,10,11]. Low-operating temperature also reduces the required refresh power and switching energy [10]. In the late 1980s, IBM first demonstrated low temperature (85 K) 512 Kbit [27,28] and 4 Mb [29] DRAMs with a significant improvement in speed and retention time compared to the DRAMs at room temperature. High carrier mobility at lower temperatures led to significant improvement in the performance of these DRAMs at cryogenic temperatures.

Fast forward to 2017, Tannu *et. al.* [8] developed an experimental setup to characterize the high-density DRAMs for quantum computing applications. They examined DRAM chips over a temperature range of 80 K to 160 K. By testing 55 DIMMS (with 750 DRAM chips) from six different vendors, Tannu *et. al.* [8] concluded that a significant number of the chips continues to work perfectly at cryogenic temperatures. Also, the error patterns (mainly transient error and permanent error) observed at cryogenic temperatures are uncorrelated errors that can be solved with conventional correction schemes like forward error correction (Chipkill and sparing [8]).

Consequently, Ware *et al.* investigated the feasibility of using 77 K DRAMs in quantum computing systems [10] and reported that DRAMs work well at 77 K without any functional errors. However, the major challenge for 77 K DRAMs is the requirement of a link interface with low thermal conductivity (a flexible cable [30]) to connect to the control processor placed at 4 K.

Another version of the CMOS DRAM is the 3-transistor (3T) DRAM cell [see Fig. 2(b) for the schematic], where two additional transistors are used for readout. The advantage of 3T DRAM cell over 1T1C DRAM is that 3T DRAM requires simpler peripheral circuitry for write/read [31]. Yoshikawa *et al.* [26] characterized 3T DRAM cells at 4 K for their hybrid Josephson-CMOS memory implementation (discussed in Section 3.C). This work reported a 40% improvement in speed and a 30% reduction in power consumption at 4 K compared to 300 K.

We summarize the reports on cryogenic DRAMs (with capacitor) in Fig. 2(c), highlighting their operating temperature and corresponding performance improvements.

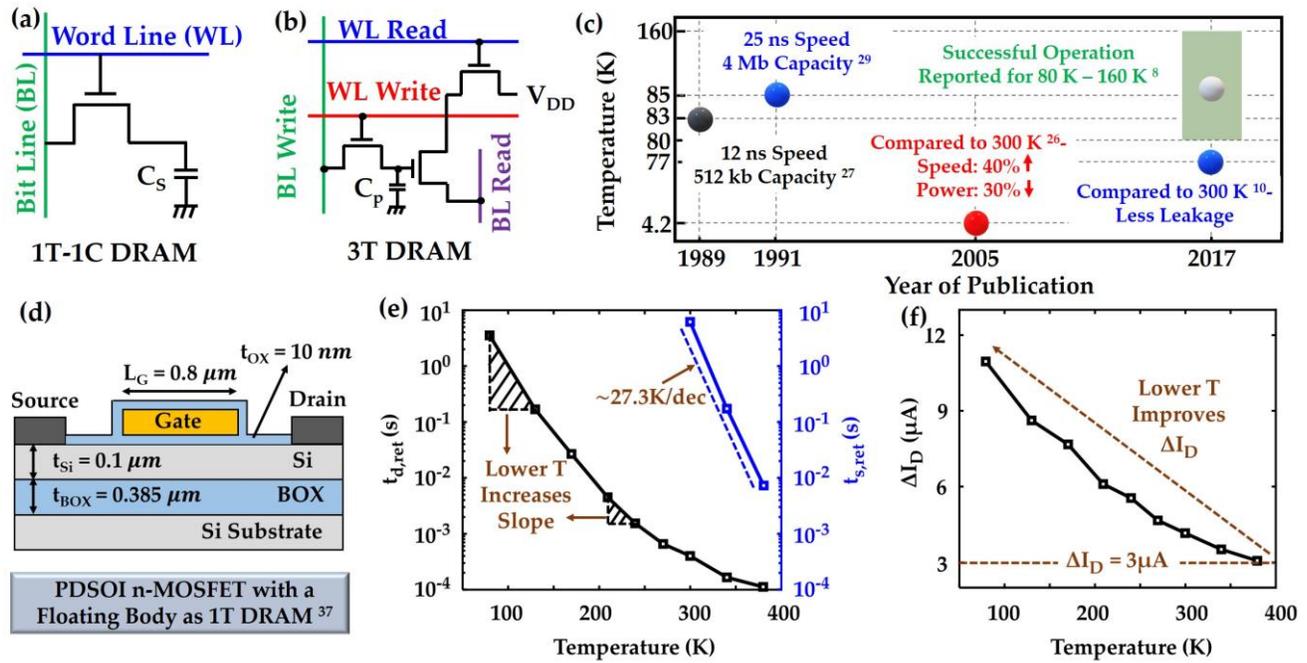

**Fig. 2:** Schematics of **(a)** 1T1C DRAM , **(b)** 3T DRAM cells. In 1T1C DRAM, the capacitor $C_s$ stores the data whereas in 3T DRAM, the parasitic capacitance stores the data. **(c)** Major reports on cryogenic DRAMs over the years. The reported effects of cryogenic temperature on different performance metrics are also mentioned. **(d)** Schematic of a PDSOI n-MOSFET with floating body [37] that can act as a capacitorless DRAM. Temperature dependence of **(e)** static retention time, $t_{s,ret}$ and dynamic retention time, $t_{d,ret}$, and **(f)** current sense margin, $\Delta I_D$. At lower temperature, the retention time and sense margin significantly improve.

### 3.A.1.2. Capacitorless 1-Transistor (1T) DRAMs

Thanks to the highly scalable structure compared to the 1T1C DRAM cells [32–36], capacitorless single transistor DRAM (1T-DRAM) devices have been explored to construct high-density memory systems. Going towards low temperatures improves the performance of the 1T DRAM cells by reducing the generation and recombination of carriers [33], a major retention failure mechanism of the 1T DRAM cells. Furthermore, an important feature of these memories is their compliance with the high capacity memory applications; this is an important requirement for quantum computer systems. These clear advantages make this memory type a strong candidate for cryogenic memory.

In 2019, a cryogenic 1T-DRAM cell (with operation over a temperature range of 380 K to 80 K) was implemented via a partially depleted silicon-on-insulator (PDSOI) n-MOSFET with a floating body [37]; see Fig. 2(d) for the schematic illustration. Memory operations (write '0', write '1', hold, and read) were demonstrated by choosing suitable bias conditions for the gate and drain of the transistor. One of the major drawbacks of the capacitorless DRAM cells is the possibility to incur retention failure. The use of low temperature significantly alleviates such failures. Figure 2(e) shows the temperature dependence of the static retention time, $t_{s,ret}$ (during hold operation) and dynamic retention time, $t_{d,ret}$ (during read operation), indicating significant improvement at lower temperatures. Compared to 300 K, $10^4$ times longer dynamic retention time, $t_{d,ret}$ [Fig. 2(e)] and 2.5 times larger current sense margin, $\Delta I_D$ [Fig. 2(f)] are achieved at 80 K [37].

### 3.A.2. Resistance-based Memories

Resistance-based memories offer better scalability (down to nano-meter), faster speed (nano-second range switching time), lower power consumption, and refresh-free operation compared with charge-based memories [38]. Moreover, resistance-based memories offer non-volatility which implies better energy-efficiency compared to the charge-based DRAMs. In these memory cells, two resistance states [known as high resistance state (HRS) and low resistance state (LRS)] define the two memory states ('0' and '1'). Memristive, ferroelectric, spintronic, quantum anomalous Hall effect (QAHE)-based, and phase-change based memories are the members of the resistance-based memory family. Figure 3 shows the schematic structures of all the members of this family as well as their signature hysteretic characteristics containing the two resistance states required for the memory operations.

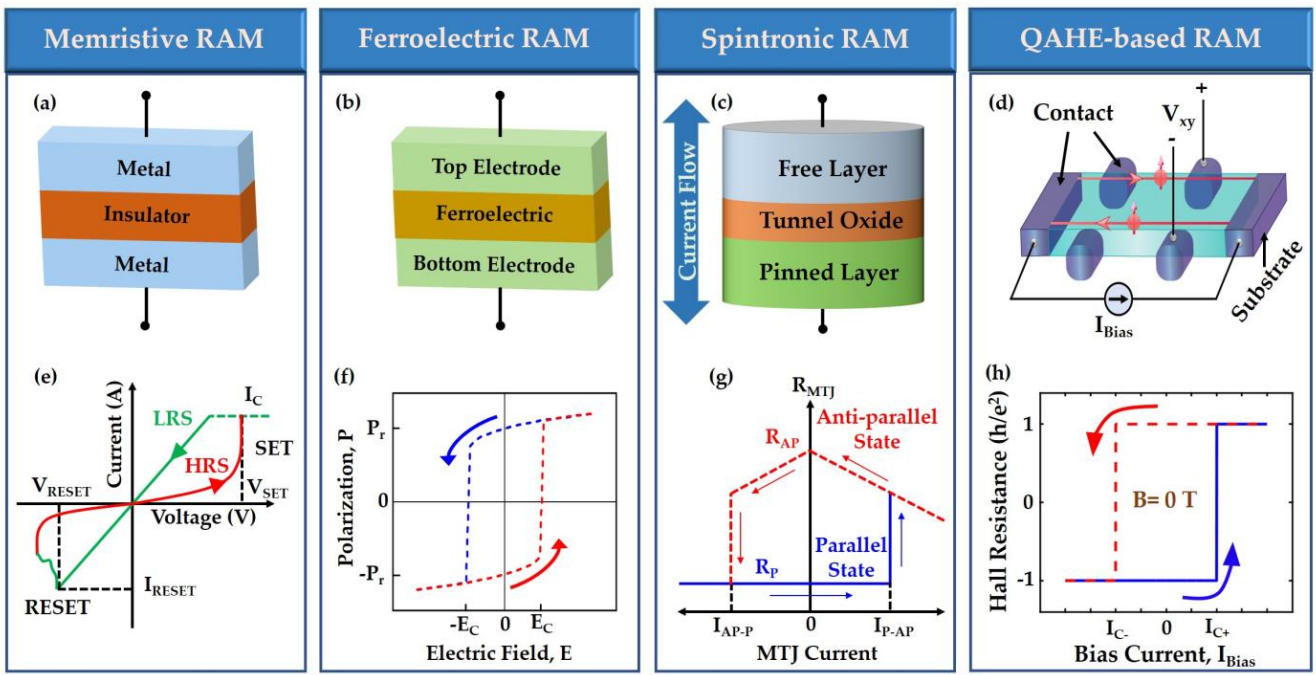

**Fig. 3:** Schematic structures of **(a)** memristive (ReRAM), **(b)** ferroelectric, **(c)** spintronic, and **(d)** QAHE-based RAM. Signature hysteresis in **(e)** the *I-V* characteristics of ReRAM, **(f)** in the *P-E* characteristics of FeRAM, **(g)** in the MTJ resistance *vs* MTJ current characteristics of MRAM, and **(h)** in the Hall resistance *vs* bias current characteristics of QAHE-based RAM. The characteristics shown in **(e)-(h)** demonstrate the two resistance states (HRS and LRS).

### *3.A.2.1. Memristive Memories*

Memristive (resistive) random-access memory (ReRAM), i.e., metal-insulator-metal (MIM) sandwiched between two electrodes [see Fig. 3(a)], is a strong candidate for cryogenic memory thanks to its sub nano-second switching time [39], low switching energy ($< 0.1$ *pJ* per bit [40]), excellent scalability (down to few nano-meters), large endurance ($10^{10}$ cycles [39]), and CMOS-compatibility. The *I-V* characteristics of these devices show a hysteretic response where two distinct resistance states (HRS and LRS) are observed [see the illustration in Fig. 3(e)]. Resistive switching has been observed in numerous material systems, such as binary transition-metal oxides (e.g., NiO [41], $TiO_2$ [42], and $HfO_2$ [43]), perovskite-type oxides [44], silicon oxides [45], and even single-molecule systems [46]. Among these, $HfO_2$-based ReRAMs show the most promise as a cryogenic memory [47–51].

Back in 2014, Ahn *et al.* analyzed the *I-V* characteristics of $Pt/Al_2O_3/HfO_x/Er/Pt$ ReRAM device over the temperature range of 40 K to 350 K [48] and found that LRS and HRS become more resistive as the temperature is lowered. Intriguingly, the device in LRS has a greater dependence on temperature than the one in HRS. Later, Shang *et. al.* [49] demonstrated the memory operation of ITO (indium-tin-oxide)/$HfO_x$/ITO MIM structure with a large $R_{OFF}/R_{ON}$ ratio, excellent endurance ($5 \times 10^7$ cycles), extrapolated retention ability (over $10^6$ s) and a wide working temperature range of 10 K to 490 K. Here, the ON-state resistance ($R_{ON}$) decreases and the OFF-state resistance ($R_{OFF}$) increases with the lowering of the temperature and therefore $R_{OFF}/R_{ON}$ improves at lower temperatures. But, it comes with the cost of larger set/reset voltages which increases the power requirement. Not so long after, in 2015, Fang *et. al.* [50] demonstrated a proper resistive switching for $Pt/HfO_x/TiN$ devices at ultra-low temperatures (4 K and 77 K). Here, the resistances in both LRS and HRS increase with the lowering of the temperature. Bolonkowski *et al.* [51] improved this structure and found that $TiN/Ti/HfO_2/TiN$ devices can operate between 4 K to 300 K temperatures. In contrast to the previous two efforts, here the resistive switching exhibited no significant change at lower temperatures. Considering the demonstration of proper switching between LRS and HRS in $HfO_x$ based ReRAM devices at ultra-low and low temperatures, these ReRAM devices can be further explored as a potential cryogenic memory.

### *3.A.2.2. Ferroelectric Memories*

Ferroelectric memories are among the most promising storage devices available to us. They are of two major types – (i) ferroelectric capacitor (FeCap) based 1T – 1C memory, and (ii) Ferroelectric field-effect transistor (FeFET) based 1T memory. The construction of FeCap memories is similar to ReRAMs- the only difference is that a ferroelectric layer is used instead of a dielectric/insulating layer [Fig. 3(b)], which provides the non-volatility. The signature of the FeCap is the hysteretic polarization (*P*) vs electric field (*E*) curve shown in Fig. 3(f). The advantages of FeCap memories include fast speed (nano-second range switching time), large endurance (about $10^{10}$ to $10^{14}$ cycles), and retention time of more than 10 years [52]. However, the destructive read operation has been one of the major issues for FeCap memories.

FeFET memories avoid the destructive read by providing separate read/write paths along with fast (~nano-second) program/erase time [53–56]. Therefore, the properties of the ferroelectric materials (an integral part of FeFET) have been extensively studied over the last few decades down to the milli-Kelvin temperature range [57–60]. Very recently, the cryogenic characteristics of an n-type FeFET containing a silicon doped hafnium oxide (Si: $HfO_2$) as a ferroelectric layer have been studied down to 6.9 K temperature [60]. This work reports that the lowering in temperature leads to an increase in the memory window with the cost of an increase in the program/erase voltage.

To explore the feasibility of realizing cryogenic FeCap and FeFET memories, several ferroelectric materials have been carefully characterized at low temperature. Examples - (i) $SrTiO_3$, oxygen-18 substituted $SrTiO_3$ and $KTaO_3$ up to 50 K [57], (ii) $PbZr_{0.5}Ti_{0.5}O_3$ thin films up to 4 K [58], and (iii) antiferroelectric zirconia up to 50 mK [59]. With the lowering of temperature, the coercive field increased for $PbZr_{0.5}Ti_{0.5}O_3$ thin films and decreased for antiferroelectric zirconia. Also, for $PbZr_{0.5}Ti_{0.5}O_3$ thin films, both the saturation and remnant polarizations kept increasing at low temperatures. These studies indicate that necessity of design trade-offs in ferroelectric memories at cryogenic temperature.

*3.A.2.3. Spintronic Memories*

Spintronic memories [magnetic random-access memories (MRAM)] have the potential to outperform the CMOS charge-based memories thanks to its non-volatile nature, high density, fast speed (switching time of < 1 ns) and low power operation, high endurance (> $10^{15}$ cycles), and long retention time (~10 years) [61]. The basic building block of MRAMs is a Magnetic tunnel junction (MTJ), consisting of two ferromagnetic materials (free layer, FL and pinned layer, PL) separated by a thin insulating layer (barrier oxide) [Fig. 3(c)]. For parallel (P) and anti-parallel (AP) magnetization in the two magnetic layers (FL and PL), two levels of magnetoresistance are observed in the MTJ, which are used to define the memory states [see Fig. 3 (g)].

Different spintronic memory devices [62–64] have been explored at cryogenic conditions to use in cryogenic MRAMs. Lang *et al.* [62] demonstrated a functioning CoFeB/MgO-based MTJ device with perpendicular magnetic anisotropy at 9 K temperature. They found a reliable and low error rate (<$10^{-4}$) switching at 9 K. Also, with the lowering of temperature to 9 K, the endurance (over $10^{12}$ cycles) improves by around three orders of magnitude. CoFeB based orthogonal spin transfer device has also been studied at 4 K temperature as a cryogenic memory element [63]. This work has shown high-speed switching (around 200 ps) and lower error rate (as low as $10^{-5}$) compared with their room temperature operations. However, this device achieves such low write error rate only for a limited pulse condition, and provides low magnetoresistance. In 2017, Yau *et al.* [64] characterized a toggle MRAM at 4.2 K for implementing their proposed hybrid JJ-MRAM memory system (discussed in Section 3.C). Along with the successful operation of the toggle MRAM at 4.2 K, this work reported an enhanced magnetoresistance at 4.2 K which improves the signal-to-noise ratio for their memory system.

*3.A.2.4. Other resistance-based Memories*

Along with the traditional resistance-based memories, exotic quantum phenomena can be leveraged to construct cryogenic memories. For example, very recently, a cryogenic non-volatile memory has been proposed based on quantum anomalous Hall effect (QAHE) [65]. QAHE is the precise quantization of Hall resistance at $\pm h/e^2$ (h = Planck's constant, e = charge of an electron) without an external magnetic field. QAHE based memory utilizes these quantized Hall resistance states ($\pm h/e^2$) observed in the twisted bilayer graphene (tBLG) moiré heterostructure [66] to define the memory states. The schematic of the tBLG heterostructure and the hysteretic behavior of Hall resistance as a function of the bias current are shown in Figs. 3(d) and (h), respectively. This design allows the user to write and read the memory cell by applying nano-ampere level bias currents which eventually makes this design very attractive as an ultra-low-power and highly scalable cryogenic memory.

Phase-change random-access memory (PCRAM) is another emerging non-volatile memory. PCRAMs have been explored for storage applications because of their non-volatility, fast speed, and superb scalability [67,68]. They utilize two phases of chalcogenide materials: the amorphous phase provides high resistance (HRS) and the crystal phase provides low resistance (LRS). A unique hysteretic behavior in resistance as a function of the applied electric field emerge in (La, Pr, Ca) $MnO_3$ up to 10 K [69]. These low temperature and stable resistive states can be switched repeatedly with the application of various voltage pulses and hence, the reported behaviors can be utilized for implementing a fast, non-volatile, and scalable cryogenic phase-change memory.

**Table I: Summary of Resistance-based Cryogenic Memories.**

| Type of Memories | Material System | Temperature Range | Effects of Lower Temperature (Compared to 300 K) | |
|---|---|---|---|---|
| Memristive | Pt/Al$_2$O$_3$/HfOx/Er/Pt [48] | 40 K – 350 K | - $R_{ON}$: $5.5 \times$; $R_{OFF}$: $0.75 \times$; $R_{OFF}/R_{ON}$: $0.3 \times$ | @ 40 K |
| | ITO/HfOx/ITO [49] | 10 K – 490 K | - $R_{ON}$: $0.75 \times$; $R_{OFF}$: $2.65 \times$; $R_{OFF}/R_{ON}$: $3.5 \times$<br>- Excellent endurance ($5 \times 10^7$ cycles)<br>- Long retention time (over $10^6$ s) | @ 10 K |
| | Pt/HfOx/TiN [50] | 4 K and 77 K | - $R_{ON}$: $2.5 \times$; $R_{OFF}$: $2 \times$; $R_{OFF}/R_{ON}$: $0.8 \times$<br>- Set voltage: $1.32 \times$<br>- Reset voltage: $1.08 \times$ | @ 4 K |
| | TiN/Ti/HfO$_2$/TiN [51] | 4 K – 300 K | - $R_{ON}$ decreases with the lowering of temperature. | @ 4 K |
| Ferroelectric | PbZr$_{0.5}$Ti$_{0.5}$O$_3$ thin films [58] | Down to 4 K | - $P_S$: $1.04 \times$; $P_R$: $1.24 \times$; $E_C$: $3.7 \times$ | @ 4 K |
| | Antiferroelectric Zirconia [59] | Down to 50 mK | - Lower critical field (in P-E Curve)<br>- High endurance (over $10^7$ cycles) | @ 50 mK |
| | Si: HfO$_2$-based FeFET [60] | Down to 6.9 K | - Memory window: $1.7 \times$<br>- Forward subthreshold slope: $0.15 \times$<br>- Increased program/erase voltage | @ 6.9 K |
| Spintronic | CoFeB/MgO-based pMTJ [62] | Down to 9 K | - $1.7 \times$ magneto-resistance<br>- 33%-93% larger switching voltage<br>- $1000 \times$ endurance ($10^{12}$ cycles)<br>- Reliable error rate ($< 10^{-4}$) | @ 9 K |
| | CoFeB-based OST device [63] | Down to 4 K | - Suppressed magneto-resistance<br>- High speed (~200 ps)<br>- Low error rate ($10^{-5}$) | @ 4 K |
| | Toggle MRAM [64] | Down to 4.2 K | - $1.5 \times$ magneto-resistance<br>- Improved signal to noise ratio | @ 4.2 K |
| QAHE-based | tBLG moiré heterostructure [65] | Down to 2 K | - Does not work above 9 K<br>- Hall resistance states become more robust at lower temperature<br>- Ultra-low Power Switching<br>- Topologically protected Hall resistance states | @ 2 K |
| PCRAM | Perovskite (La, Pr, Ca)MnO$_3$ [69] | Down to 2 K | - High distinguishability ($10^5$) between two phase resistances | |

Table I summarizes the attempts on cryogenic resistive-based memories along with the effects of cryogenic temperature on their operation and performance.

### 3.B. Superconducting Memories

Although the non-superconducting memories (discussed in Section 3.A.) can offer excellent scalability, these memories have higher power demand and provide lower speed compared to the superconducting SFQ circuits and systems. JJ and SQUID-based SFQ technology has emerged as a promising platform to design the control processor for quantum computers. SFQ processors can operate at ultra-low temperature (4 K) with high speed (close to THz) and low switching energy (0.1-1 atto-Joule) [8]. Therefore, superconducting memory is an obvious choice for cryogenic applications. However, superconductor-insulator-superconductor (SIS) JJ and SQUID-based memories offer unreasonably low storage capacity due to (i) the large cell area (a few hundreds of $\mu$m$^2$), (ii) transformer-based coupling with address lines, and (iii) flux trapping [70–80]. Trapping of magnetic flux is one of the major issues for the superconducting circuits that adversely affects the performance [81–83]. Typically, the trapping of magnetic flux occurs during the cooling process when the metal films transform into the superconducting state. The trapped fluxes affect the superconducting circuits in two ways- (i) can either penetrate JJs or (ii) can couple with magneto-sensitive gates [81]. The first step to prevent flux trapping is to shield of the earth magnetic field which cannot solve the problem completely. Therefore, it is necessary to keep some space on the superconducting chips so that frozen vortices remain far from the magneto-sensitive portions of the chips [82]. Now, most of the JJ-based memories use address lines couple with transformers which are magneto-sensitive. Therefore, this flux trapping imposes a limitation on the integration density of JJ and SQUID-based memories. Besides the storage capacity, energy-efficiency at large-scale is an issue [84,85]. As Holmes *et al.* [84] demonstrated that the available superconducting memories consume too much power and therefore, they cannot be used as memory at 4 K temperature. The fabrication of scaled JJs is a major step towards implementing high density SFQ memories. Ref. [86] provides a detailed review of the development in the fabrication technology of JJs. Tolpygo *et al.* [87] have developed a fabrication process that can yield circuits with over $7 \times 10^4$ JJs on a $5 \times 5$ mm$^2$ chip. This ongoing development in the fabrication technology of JJs may lead to scalable JJ-based memory. Another popular approach to attain high capacity is to pursue the hybrid superconductor-semiconductor schemes combining CMOS technology and Josephson junction-based superconducting technology. This hybrid approach brings in some unique challenges which we will discuss later (section 3.C.). Two feasible alternatives can be the magnetic JJ (MJJ) and superconducting memristor-based memories, which have the potential to offer high capacity, speed, and energy efficiency while being capable of integration on a single chip with SIS JJs. These traits result in a fast operation, even at a similar clock speed with the fast SFQ control processor and SFQ digital circuits. It is worthwhile to elaborate on superconducting memories in three subsections: (i) SIS JJ-based, (ii) magnetic JJ-based, (iii) superconducting memristor-based, and (iv) ferroelectric SQUID-based memories.

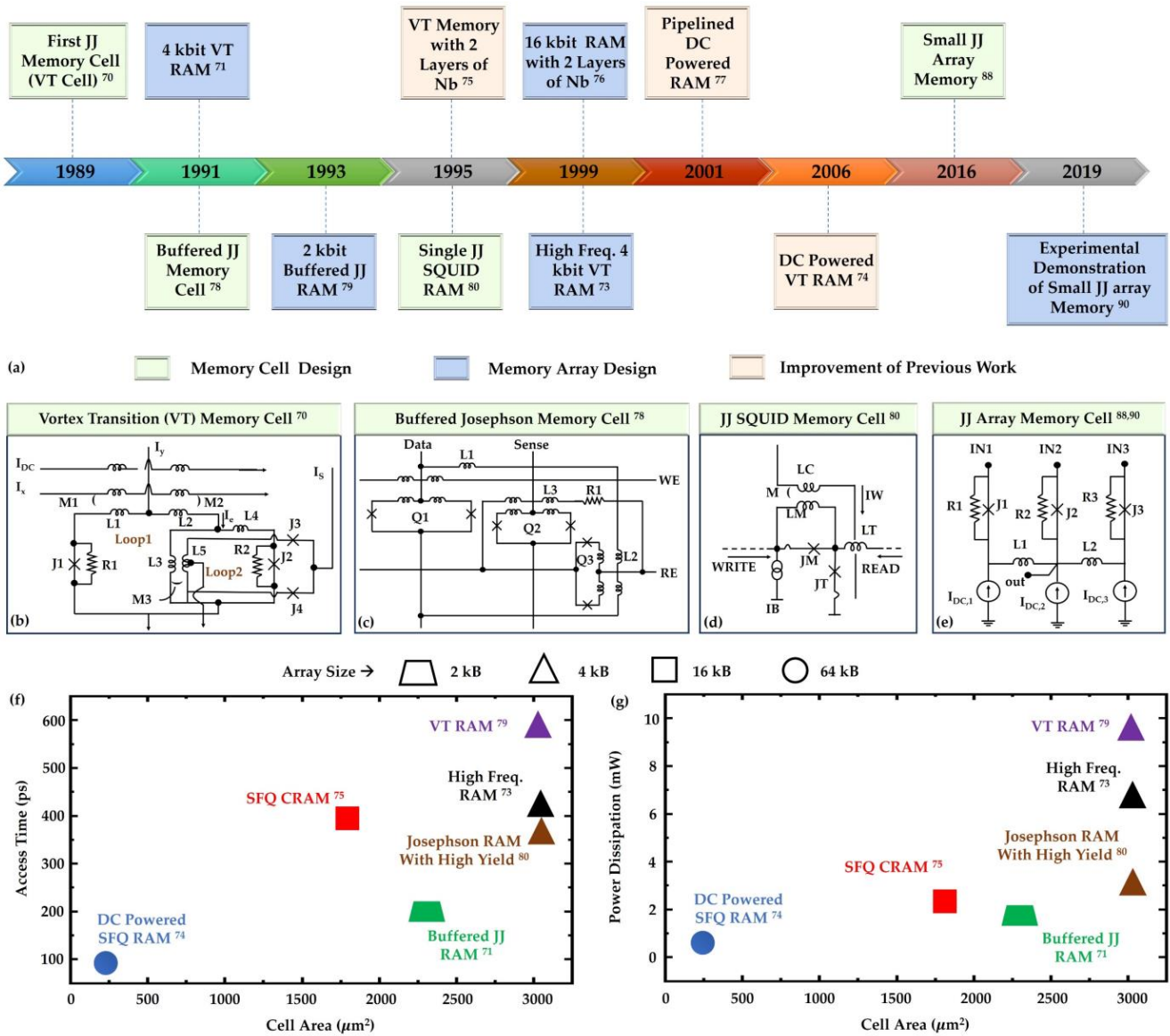

**Fig. 4:** **(a)** Timeline demonstrating the enormous research attempts on SIS JJ based memories over the last few decades. Schematics of **(b)** Josephson vortex transition memory cell [70], **(c)** buffered Josephson memory cell [78], **(d)** single JJ SQUID based memory cell [80], and **(e)** three inductively coupled SIS JJs based memory cell [86,88]. **(f)-(g)** Comparison of SIS JJ-based memories based on important performance metrics.

*3.B.1. SIS JJ-based Memories*

SIS JJ-based cryogenic memory design has garnered enormous research attention over the last few decades [70-80] [summarized in the timeline of Fig. 4 (a)]. Despite all the efforts, these designs suffer from low capacity. To date, only 4 kbit memory has been demonstrated experimentally [73]. Tahara *et. al.* [70] first demonstrated a nondestructive read-out (NDRO) Josephson vortex transition (VT) memory cell in 1989. The memory cell consists of two superconducting loops (contains Nb/AlO$_x$/Nb JJs and inductors) and a two-junction interferometer gate as a sense gate [schematic shown in Fig. 4(b)]. In Loop 1 and Loop 2 [Fig. 4(b)], resistors are connected in parallel to JJs ($J_1$ and $J_2$) to ensure suitable damping conditions for the junctions. In this design, Loop 1 stores the information in form of an SFQ pulse, and the stored data is read by the switching of the sense gate caused by the vortex transition [70] in Loop 2, which again depends on the stored SFQ pulse in Loop 1. In a later work, two 4-kbit RAMs, composed of 64×64 [71] and 256×16 [73] VT memory cells have been demonstrated. Nagasawa *et al.* [74] later modified the VT memory cell and proposed a pipeline structure to design a DC-powered RAM. To control the modified VT memory cell with DC signals, a two-junction SQUID gate is used as the write gate and a control line is magnetically coupled to Loop 1. In a subsequent work, the VT memory cell was further improved by reducing the number of Nb layers from three to two [75]. Using this improved design, a 16 kbit RAM comprising four blocks (4 kbit each) has been demonstrated in Ref. [76]. A pipelined DC-powered RAM structure had also been proposed later for this 16 kbit Memory [77]. Yuh *et. al.* [78] demonstrated SQUID-based NDRO memory cell [shown in Fig. 4(c)] which also consists of Nb/AlO$_x$/Nb JJs and inductors like the VT memory cell and later implemented a 2 kbit RAM using these cells [79]. This design uses three sense gates (unlike just one in VT memory cells) to solve the 'half select' problem of SIS JJ-based memories.

In 1995, Polonsky *et al.* [80] proposed a new design concept for the JJ-based RAMs shown in Fig. 4(d). The designed memory cell consists of single-junction SQUIDs which are serially connected to the bit lines and coupled inductively to the word lines. Write and read operations are performed by sending pulses in appropriate directions into bit lines and DC pulses with appropriate polarity along word lines.

Recently in 2016, Nair *et al.* [88,89] developed a new design for SIS JJ-based memory cell shown in Fig. 4(e) and experimentally demonstrated the memory operation in 2019 [90]. The memory cell consists of three inductively coupled SIS JJs and offers three memory states. Write operations are performed by applying SFQ pulses into appropriate junctions and read operation is performed using one of the write mechanisms. However, based on the mechanism used to sense the stored data, one of the read operations (1/0) will be destructive (half-destructive read). However, non-destructive readout can be achieved with the same design and same readout technique; only by sending a weaker pulse compared to the ones sent during the write operations [91,92].

Next, we compare the performances of the above-mentioned SIS JJ-based memories. Cell area, speed, storage capacity, and energy efficiency are the major performance metrics for cryogenic memories. Figures 4 (f) and (g) compare the cell area, access time, capacity, and power consumption for the notable SIS JJ-based memories. Ultra-high-speed and low power consumption are attractive features, but scalability is a major concern for these memories. More detailed reviews of this memory technology are available in Refs. [93,94].

*3.B.2. Magnetic JJ-based Memories*

SIS JJ is the main building block of superconducting electronics and the control processor of superconducting qubit-based quantum computers. Therefore, a memory that is compatible with the speed, power, and fabrication method of SIS JJs should be a strong candidate for cryogenic memory applications. However, SIS JJ-based memories suffer from low capacity. As an alternative, in the 1990s, the idea of combining ferromagnetic and superconducting materials to develop a high-capacity cryogenic memory was explored [95]. Magnetic JJ (MJJ) is a version of Josephson junction, built by sandwiching a ferromagnetic layer between two superconducting materials [schematic shown in Fig. 5(a)].

MJJ and SIS JJ have the same fabrication process and hence, MJJs can be integrated on a single chip with SIS JJs [96]. The most significant advantage of MJJs is their ability to be in a Josephson state with the inversion of phase difference which is known as π-state [97,98]. The incorporation of the ferromagnetic layer in the Josephson junction creates a hysteretic dependence of the critical current on the external magnetic field [see Fig. 5(b)]. A small applied magnetic field affects the magnetization of the ferromagnetic layer which eventually changes the $I_C$ of the junction. This change in $I_C$ allows two distinct states to be realized with high (logic '0') and low (logic '1') $I_C$ [marked as $I_{C0}$ and $I_{C1}$, respectively in Fig. 5(b)]. A small magnetic field, that can be achieved by a small control current, can perform the non-volatile change of $I_C$ of the MJJ based-memories. MJJs are one of the strongest candidates for cryogenic memory in quantum computers and SFQ systems, thanks to their high storage capacity and compatibility with the SIS JJs.

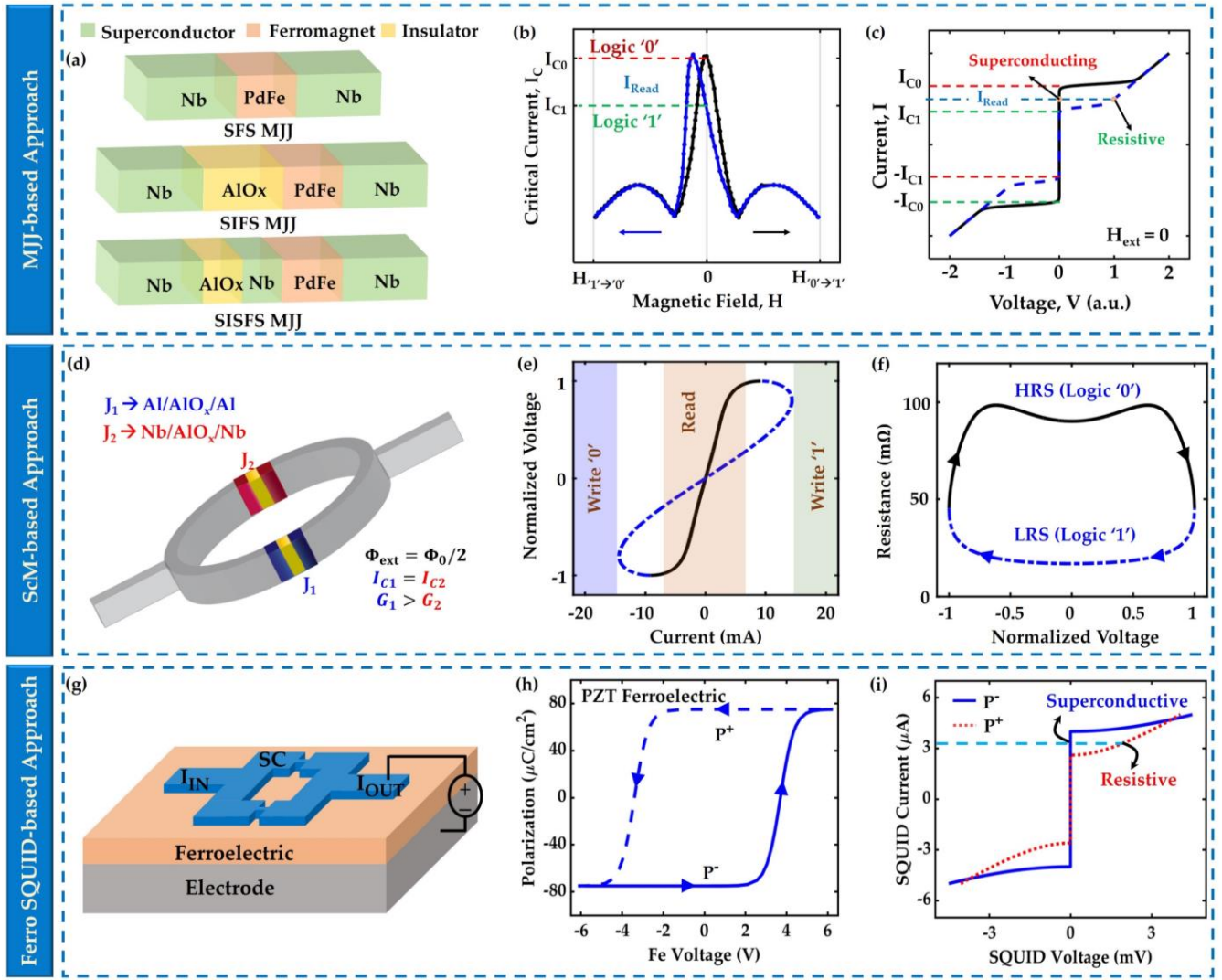

**Fig. 5:** **(a)** Schematic of SFS MJJ, SIFS MJJ, and SI(S)FS MJJ structures. **(b)** Hysteretic dependence of critical current $I_C$ on magnetic field [99-101]. Magnetic field sweep directions are shown by arrows. **(c)** Current-voltage characteristics of MJJs. **(d)** Schematic of a CA-SQUID device [118], consisting of two SIS JJs with different superconducting electrodes. Two JJs have the same critical current but different conductance. The application of a magnetic flux of $\Phi_0/2$ suppresses the effect of equal critical current but not the effect of phase-dependent current component, due to the asymmetry in conductance. **(e)** I-V characteristics and **(f)** resistance vs voltage characteristics of the CA-SQUID-based superconducting memristor [118]. Two resistance levels [HRS (high resistance state) and LRS (low resistance state)] are utilized to define the memory states. Regions are marked in the I-V characteristics for different memory operations (write and read). **(g)** Device structure of a ferroelectric SQUID [123]. **(h)** Polarization vs voltage characteristics and **(i)** current vs voltage characteristics for the device. Two polarization states (positive and negative) of the ferroelectric material affect the crictical current of the SQUID. This hysteretic dependence of critical current on the polarization of the ferroelectric is utilized in this voltage-controlled non-volatile memory [123].

Ryazanov *et al.* [96] first proposed the MJJ-based memory with fast and energy-efficient SFQ switching readout. They demonstrated the ability of a superconductor-ferromagnet-superconductor (SFS) JJ (Nb/Pd$_{0.99}$Fe$_{0.01}$/Nb) to operate as a Josephson magnetic switch. They utilized the hysteretic behavior of $I_C$ on the external magnetic field in their memory design. A similar hysteretic behavior has been observed in SFS JJs with both strong and weak ferromagnets [99–103]. The speed of an MJJ-based memory cell depends on the inductance of the control current line and the intrinsic switching time ($\tau_J$) of the corresponding SFS junction. $\tau_J$ is given by [96]:

$$\tau_J = \frac{\Phi_0}{2\pi I_C R_N} \ , \tag{1}$$

Where, $\Phi_0$ is the single flux quantum and $R_N$ is the normal resistance of the junction [72]. SFS JJs provide a low junction characteristic voltage $V_C$ (= $I_C R_N$) (in the order of nano volts) [96]. It limits the switching time to ~ 100 ns which is not compatible with the speed of the SIS JJs. To increase $V_C$, Ryazanov et al. [96] proposed to insert an additional insulator layer in the SFS Josephson junction [Fig. 5(a)]. A superconductor-insulator-ferromagnetic-superconductor (SIFS) MJJ (Nb/Al-AlO$_x$/Pd$_{0.99}$Fe$_{0.01}$/Nb) with a higher $V_C$ has also been shown. This design retains the magnetic memory properties. Later, both weak and strong ferromagnets have been used to design SIFS MJJs [104,105]. The characteristic voltage had been further increased by inserting an additional superconducting layer [SI(S)FS JJ] [106–108]. Using SI(S)FS JJ [Fig. 5(a)], a maximum $V_C$ of ~ 700 $\mu V$ (which implies ~ 200 GHz switching) was obtained [107].

Besides the MJJs with one ferromagnetic layer, multiple ferromagnetic layers have been used in MJJs (S/F$_1$/F$_2$/.../S). These MJJs are commonly known as spin-valve JJs. Such MJJs allow tuning of the ground-state phase difference and critical current by the mutual orientation of the ferromagnetic layers [109–112]. Bell et al. first demonstrated spin-valve JJ-based cryogenic memory[113]. Similar spin-valve JJs had been demonstrated with several ferromagnetic materials and alloys, such as - CuNi [104], PdNi [114], PdFe [96,102], Ni [98,115], Fe [99], Co [99], NiFe [99,116], NiFeNb [117], NiFeMo [100], and NiFeCo [116].

Quantized Abrikosov vortex memory is another example of MJJ-based memory which was demonstrated using two different architechtures- Josephson spin-valve structures and planar JJs [118]. Two resistance states (high and low) are observed in these memory cells depending on the presence and absence of the *Abrikosov vortex* [118]. The vortex can be introduced (removed) by applying a positive (negative) current of ∼20 $\mu A$. These memory cells demand an extremely low write energy (in the order of atto-Joules), offer non-volatility (due to the quantized nature of the *Abrikosov vortex*), and can be scaled to nano-meter size. However, the challenge is that the required magnetic field becomes significantly higher (~1 kOe) for smaller structures [118].

Finding a suitable ferromagnetic material for MJJs is challenging because strong and weak ferromagnets used in MJJs suffer from short characteristic length and severe depression of critical current, respectively. Weak ferromagnets have two key advantages over strong ferromagnetic materials. The first advantage is from the magnetic side- weak ferromagnets require less switching energy. Stoner-Wohlfarth theory [119] predicts that the switching field required by a single-domain nanomagnet is proportional to the magnetization, and the switching energy is proportional to the square of its magnetization. Therefore, lower magnetization implies lower switching energy. The second advantage is from the superconducting perspective. Strong ferromagnets have a short characteristic length scale ($\xi_F$) [exceeding this length will lead to oscillatory switching between 0/$\pi$-type JJs] [99]. Strong ferromagnetic materials (such as Co or Fe) typically exhibit $\xi_F < 1$ nm [99], which implies that the junction properties will fluctuate even if the average thickness of the ferromagnetic layer changes by a fraction of the atomic monolayer. On the other hand, weak ferromagnets show much larger $\xi_F$ which makes it easier to control the thickness of the ferromagnets and hence, junction properties show a smaller sample-to-sample variations [97,116]. Despite these two strong advantages, weak ferromagnets are not yet prominent in spin-valve JJ based memory, as they lead to drastic reduction in critical current [97]. Low critical current in MJJs will lead to a smaller hysteresis window in the characteristics shown in Fig. 5(b), making it difficult to sense the memory states.

Another challenge for MJJ-based memory cells is the array design. To write into the memory cells, a suitable magnetic field is applied (possibly using an on-chip magnet). This magnetic field may interfere with the stored data in the neighboring cells. Another option is to utilize a current-carrying coil to apply the magnetic field [120]. The usage of current-carrying coil, however, leads to slow switching (~milli-seconds) [120]. Therefore, the design of a large array based on MJJ remains an unsolved challenge.

### 3.B.3. Superconducting Memristor-based Memory

One of the most recent designs of cryogenic memory utilizes a *superconducting memristor* (ScM) [122] as the storage element. The ScM can be realized using a conductance asymmetric SQUID (CA-SQUID), harnessing the phase-dependent conductance of JJs. The ScM exhibits a pinched hysteresis loop in its current-voltage characteristics (reminiscent of an ideal memristor), and combines the scalability of classical memristors with the ultrafast speed and high energy efficiency of JJs [122,123]. Figure 5(d) shows the schematic of a CA-SQUID-based superconducting memristor, where two SIS JJs are connected in parallel. Traditionally, the two JJs in CA-SQUID are designed with two different superconducting electrodes to ensure same critical current and asymmetric conductance.

The dynamics of CA-SQUID can be explained with the modified resistively and capacitively shunted junction (RCSJ) model [124,125] with four shunt paths- (i) the Josephson inductance to capture the supercurrent [126], (ii) a constant resistance ($R_N$) to capture the single-electron tunneling [127], (iii) a capacitance ($C_J$) due to the junction capacitance, and (iv) a dissipative current path due to the phase-dependent conductance of JJ. Now, with a suitable flux bias ($\Phi = \Phi_0/2$), the effect of critical current can be suppressed while maintaining the asymmetric conductance; it leads to the pinched hysteresis loop in the *I-V* characteristics of the CA-SQUID [shown in Fig. 5(e)] [122]. The two resistance states of the CA-SQUID are used to define the memory states [Fig. 5(f)]. Array level implementations of CA-SQUID is on the horizon. In fact, a recent work [123] reports a cryogenic memory array design which uses a superconducting memristor-based memory cell and heater cryotron (*hTron*) [121]-based access device. Experimental implementation of this novel design concept awaits.

### 3.B.4. Ferroelectric-SQUID-based Memory

Ferroelectric-SQUID (FeSQUID) is a hybrid superconductor-ferroelectric device where, a SQUID containing two parallel weak links is fabricated on top of a ferroelectric substrate [128]. Fig. 5(g) shows the schematic of the FeSQUID device. The ferroelectric layer has a switchable polarization that can be controlled by a voltage bias. Polarization vs. voltage hysteretic characteristics of a Lead Zirconate Titanate (PZT) ferroelectric material is shown in Fig. 5(h). As seen in Fig. 5(h), the ferroelctric material can show two voltage-controlled non-volatile polarization states ($P^-$ and $P^+$). These two states can be utilized to define the two memory states ('0' and '1', respectively). Therefore, the write operation of this type of memory is voltage-based. Now, the polarization state of the ferroelectric material determines the bound charge in the ferroelectric which eventually controls the superconductor charge density and related parameters. One of the most important parameters of the SQUID on top of the ferroelectric substrate that changes based on the polarization is the critical temperature. As the critical temperature consequently affects the critical current of the SQUID, $P^-$ and $P^+$ polarization states result in two distinct level of critical currents ($I_{C,high}$ and $I_{C,low}$). Negative polarization state ($P^-$) leads to higher critical current than the positive polarization state ($P^+$) as shown in Fig. 5(i). Now, for read operation, a suitable bias current ($I_{Read}$) can be applied to the FeSQUID to invoke superconducting/non-superconducting behavior based on the stored memory states. For $|I_{C,low}| < |I_{Read}| < |I_{C,high}|$, as seen in Fig. 5(i), FeSQUID shows superconducting (0 V) behavior for $P^-$ polarization state and non-superconducting behavior (nonzero voltage) for $P^+$ polarization state. The difference in the voltage across the FeSQUID device based on the polarization state during read operation can be utilized for the sensing mechanism. FeSQUID-based memory combines the advantages of the superconducting devices (such as high speed and low power consumption) with the advantages of the ferroelectric materials (such as non-volatility, scalability, voltage-based control, and separate read-write paths).

### *3.C. Hybrid Memories*

Conventional non-superconducting memories provide excellent scalability but suffer from lower speed and higher power demand compared to the JJ-based control processor. SIS JJ-based superconducting memories, on the other hand, suffer from very low capacity [73,80]. To combine the advantages of these two technologies, hybrid memories were proposed. The hybrid design, which utilizes the best features of each technology, is a promising approach to develop high speed, energy-efficient, and high capacity cryogenic memories [129]. In this approach, highly scalable memories (such as DRAMs, MRAMs, etc.) are used as the storage element, and JJ-based superconductive devices are used to access them. Figure 6(a) illustrates the concept of implementing a hybrid memory system that combines the advantages of both technologies.

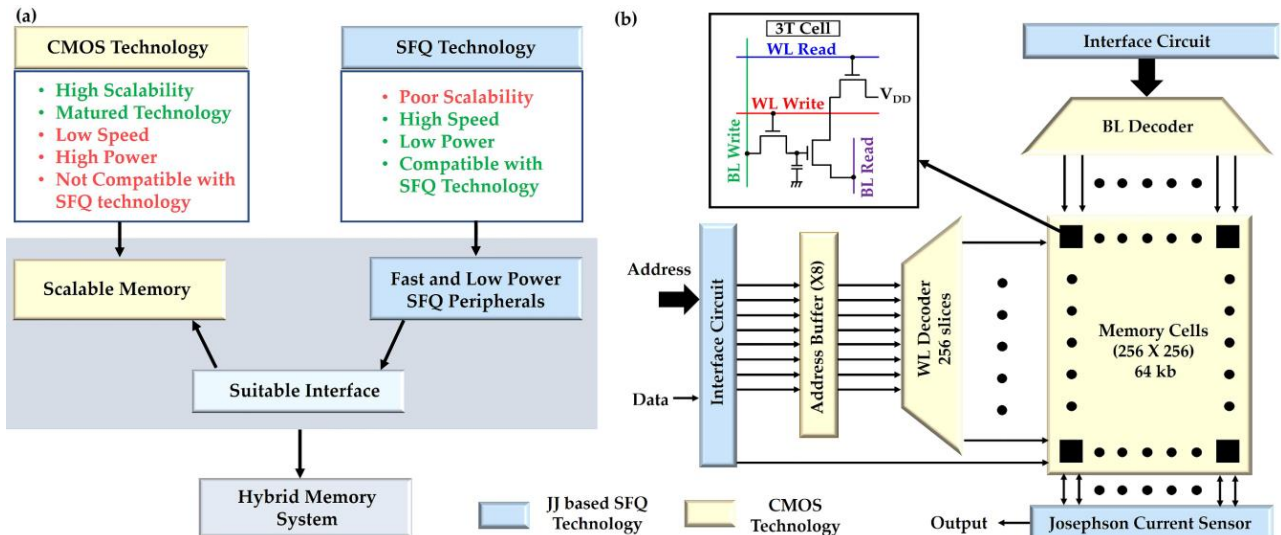

**Fig. 6:** (**a**) Illustration of a hybrid memory system, where non-supercomputing technology is used as the scalable memory system and superconducting technology is used to access the memory. (**b**) Example of a hybrid memory system for cryogenic applications proposed in Ref. [121]. Here, 3T DRAM is used as the memory element and JJ based circuits are used as the access scheme.

Ghoshal et al. [129] first put forward this concept and some simulations and measurements were reported in their subsequent works [130–133] to verify the feasibility of the idea. They used 3T DRAMs as the storage element and demonstrated a 64 kb (256 × 256) hybrid memory system. Figure 6(b) shows the high-level block diagram of their hybrid memory system. A similar hybrid memory system was also designed (for operation at 4 K) using the CMOS static RAM (SRAM) as the memory cell [134]. Moreover, in stead of using only JJs to design the access circuit for JJ-CMOS hybrid memories, three-terminal nano-cryotrons have also been used for CMOS DRAM memory systems [135].

Integration of MRAMs with Josephson junctions is another interesting way to develop a hybrid cryogenic memory (Josephson-MRAM memory) [64]. Ref. [64] first examined the operation of magnetic tunnel junction (MTJ) devices at 4.2 K, demonstrating successful switching operations. Spin Hall effect (SHE)-based MTJ devices [136] were also explored for cryogenic memory [137]. An energy-efficient (6 pico Joule per switching operation) 4 × 4 array was demonstrated [137] using an SHE-MTJ and an $hTron$. The memory cells in the array were accessed with 100 $\mu A$ current signals which makes the design compatible with SFQ control circuits.

Unlike the above-mentioned hybrid memories, where two technologies are placed at the same temperature, Mukhanov et al. [138] suggested a hybrid cryogenic memory system where two technologies were placed at different temperatures. This approach integrates a high-density semiconductor memory at room temperature and a high-speed 4.2 K rapid single flux quantum (RSFQ) cache that is integrated on a chip with an RSFQ processor. This approach is not suitable for large-scale quantum computers due to the interconnection between the room temperature memory and the cryogenic cache. Nevertheless, it is suitable for digital RF receivers due to the high capacity of semiconductor memory and the fast readout ability of superconductive devices. The hybrid memories can be one of the potential candidates for the storage system in quantum computing and other cryogenic applications.

## 4. Comparison of Cryogenic Memories: Pros, Cons, Challenges and Future Prospects

In Section 3, we discussed the basic concepts and working principles of the major variants of cryogenic memory technologies. Here we present a critical comparison among these technologies in light of the requirements for quantum computing and superconducting electronics. This comparative discussion will guide the readers to the advantages, disadvantages, and challenges faced by these cryogenic memory devices.

The non-superconducting memories such as CMOS DRAMs, memristive, spintronic, and ferroelectric memories are well matured compared to all the other memory technologies. It may take a long time for the emerging superconducting memories to catch up. Remarkably, all the non-superconducting memories offer significantly better scalability compared to the superconducting memories. In retrospect, most of the conventional memories exhibit successful operation (even with performance improvement in some cases) at 4 K, implying that these memories are suitable for integration with the superconducting control processor in terms of operating temperature. However, 4 K non-superconducting memories suffer from low-speed and high-power issues compared to the fast and energy-efficient superconducting SFQ circuits, systems, and processors. Therefore, these memories will require a CMOS-based control processor because they cannot operate at the same speed and the same power budget as the SFQ circuits [8,10]. Moreover, the non-superconducting CMOS devices consume high power at large scale that cannot be sustained at 4 K temperature.

On the other hand, SIS JJ and SQUID-based emerging superconducting memories are attractive owing to their 4 K operations and excellent compatibility with the control processor in terms of operating temperature, speed, and power. The only remaining challenge is the implementation of a high-capacity memory system for quantum computers and other applications. To solve the scalability issue, magnetic JJ-based memory and hybrid variants can be promising. However, both options have their own challenges. Magnetic JJ-based memories offer compatibility with the SIS JJ-based SFQ circuits/systems/processors on integration in the same chip, operating temperature, speed, and power. However, a suitable ferromagnetic material is yet to be found partly because they suffer from fluctuations arising from the ferromagnetic layer thickness and the severe depression of the critical current. Also, the array design for MJJ-based memories is difficult.

Another alternative for scalable cryogenic memory is the hybrid approach which is envisioned to solve the issues faced by the non-superconducting memories and the emerging superconducting technology. It combines the best features of both the technologies such as the maturity and high capacity of the non-superconducting memories and the fast and energy-efficient operation of the superconducting circuits. Since 4 K operation of the conventional non-superconducting memories has been reported, the only remaining challenge is to find a suitable electrical interface between the non-superconducting and superconducting devices. The interface circuits are mainly cryogenic amplifier circuits which amplify millivolt signals of Josephson circuits to volt level signals required for CMOS circuits. Superconducting amplifiers [139], semiconducting amplifiers [140], and hybrid superconductor-semiconductor amplifiers [129] are being explored to find suitable Josephson-CMOS interface circuits. Designing these interfaces is challenging and requires future research. Their speed and energy demand must not impose any additional bottlenecks. It is worthwhile to note, unlike other hybrid approaches, the memory based on the SHE-MTJ and *hTron* [137] does not require such an interface circuit. We believe, as long as the scalability issue of the superconducting memories persists, hybrid memories may be the most feasible and compatible options for cryogenic applications.

## 5. *Conclusion*

Quantum computers and superconducting SFQ circuits/systems are envisioned to solve the limitations and challenges of the classical computers and CMOS electronics, respectively. However, the lack of a compatible cryogenic memory system has limited them to only niche applications. In this review, we discussed the major efforts on implementing a suitable cryogenic memory for broader applications. In particular, we highlighted cryogenic characterization of non-superconducting room temperature storage devices (DRAMs, ReRAMs, MRAMs, FeRAMs, etc.), SIS JJ, MJJ, and ScM memories, as well as hybrid topologies. We discussed the major challenges faced by these memories in their integration with quantum computers and the SFQ circuits. In brief, cryogenic versions of the non-superconducting memories suffer from speed and power compatibility with the SFQ control processor even though they offer a large capacity. Conversely, JJ-based emerging memories are compatible with the control processor and SFQ circuits, but they lack scalability due to their larger size and the requirement of inductive coupling. On the other hand, hybrid memories can combine all the advantages of superconducting and non-superconducting technologies. However, the design of suitable interface circuits remains a challenge. Moreover, for a reliable superconducting and hybrid memory, the operating temperature must be kept way below the critical temperature of the superconducting material. Till date, it has not also been explored how the fluctuation in the operating temperature affects the reliability of the superconducting and hybrid memory systems. All these design conflicts and opportunities have left the field wide open for future research leveraging novel materials, devices, and architectures.

**Data availability**

The data that support the plots within this paper are available from the corresponding author on reasonable request.

**Author contributions**

All authors conceived the idea of this Review. S. A. performed the literature analysis and collected data. All authors took part in writing the manuscript, discussed the data, and contributed to the final manuscript. A. A. supervised the project.

**Competing interests**

The authors declare no competing interests.

**References**

1. Mujtaba, H. Cerebras Wafer Scale Engine Is A Massive AI Chip Featuring 2.6 Trillion Transistors & Nearly 1 Million Cores. https://wccftech.com/cerebras-unveils-7nm-wafe-scale-engine-2-largest-ai-chip-ever-built/.
2. Masanet, E., Shehabi, A., Lei, N., Smith, S. & Koomey, J. Recalibrating global data center energy-use estimates. *Science (80-. ).* **367**, 984–986 (2020).
3. Likharev, K. K. & Lukens, J. Dynamics of Josephson Junctions and Circuits. *Phys. Today* (1988) doi:10.1063/1.2811641.
4. Mukhanov, O. A. *et al.* Superconductor digital-RF receiver systems. *IEICE Trans. Electron.* (2008) doi:10.1093/ietele/e91-c.3.306.
5. Vernik, I. V. *et al.* Cryocooled wideband digital channelizing radio-frequency receiver based on low-pass ADC. in *Superconductor Science and Technology* (2007). doi:10.1088/0953-2048/20/11/S05.
6. NSA. *Superconducting Technology Assessment. National Security Agency Office of Corporate Assessments* (2005).
7. Feynman, R. P. Quantum mechanical computers. *Found. Phys.* (1986) doi:10.1007/BF01886518.
8. Tannu, S. S., Carmean, D. M. & Qureshi, M. K. Cryogenic-DRAM based memory system for scalable quantum computers: A feasibility study. in *ACM International Conference Proceeding Series* (2017). doi:10.1145/3132402.3132436.
9. Filippov, T. V. *et al.* 20 GHz operation of an asynchronous wave-pipelined RSFQ arithmetic-logic unit. in *Physics Procedia* (2012). doi:10.1016/j.phpro.2012.06.130.
10. Ware, F. *et al.* Do superconducting processors really need cryogenic memories? The case for cold DRAM. in *ACM International Conference Proceeding Series* (2017). doi:10.1145/3132402.3132424.
11. Patra, B. *et al.* Cryo-CMOS Circuits and Systems for Quantum Computing Applications. *IEEE J. Solid-State Circuits* (2018) doi:10.1109/JSSC.2017.2737549.
12. Veldhorst, M. *et al.* An addressable quantum dot qubit with fault-tolerant control-fidelity. *Nat. Nanotechnol.* (2014) doi:10.1038/nnano.2014.216.
13. Veldhorst, M. *et al.* A two-qubit logic gate in silicon. *Nature* (2015) doi:10.1038/nature15263.
14. Chow, J. M. *et al.* Implementing a strand of a scalable fault-tolerant quantum computing fabric. *Nat. Commun.* (2014) doi:10.1038/ncomms5015.
15. Dicarlo, L. *et al.* Demonstration of two-qubit algorithms with a superconducting quantum processor. *Nature* (2009) doi:10.1038/nature08121.
16. Kawakami, E. *et al.* Electrical control of a long-lived spin qubit in a Si/SiGe quantum dot. *Nat. Nanotechnol.* (2014) doi:10.1038/nnano.2014.153.
17. Hastings, M. B., Hastings, M. B., Wecker, D., Bauer, B. & Troyer, M. Improving quantum algorithms for quantum chemistry. *Quantum Inf. Comput.* (2014).
18. Shor, P. W. Polynomial-time algorithms for prime factorization and discrete logarithms on a quantum computer. *SIAM J. Comput.* (1997) doi:10.1137/S0097539795293172.
19. Kessler, M. F. The Infrared Space Observatory (ISO) mission. *Adv. Sp. Res.* (2002) doi:10.1016/S0273-1177(02)00557-4.
20. Kirschman, R. K. Low temperature electronic device operation. in *Symp. Electrochemical Society* vol. 14 (1991).
21. Dean, M., Foty, D., Saks, N., Raider, S. & Oleszel, G. Low temperature microelectronics: opportunities and challenges. in *Proc. Symp. Low Temperature Electronic Device Operation, Electrochemical Society* vol. 91 25–37 (1991).
22. Hornibrook, J. M. *et al.* Cryogenic Control Architecture for Large-Scale Quantum Computing. (2015) doi:10.1103/PhysRevApplied.3.024010.
23. Tannu, S. S., Myers, Z. A., Nair, P. J., Carmean, D. M. & Qureshi, M. K. Taming the Instruction Bandwidth of Quantum Computers via Hardware-Managed Error Correction. *Proc. 50th Annu. IEEE/ACM Int. Symp. Microarchitecture* **13**,.
24. Tuckerman, D. B. *et al.* Flexible superconducting Nb transmission lines on thin film polyimide for quantum computing applications. *Supercond. Sci. Technol.* (2016) doi:10.1088/0953-2048/29/8/084007.
25. Likharev, K. K. Superconductor digital electronics. *Phys. C Supercond. its Appl.* (2012) doi:10.1016/j.physc.2012.05.016.
26. Yoshikawa, N. *et al.* Characterization of 4 K CMOS devices and circuits for hybrid Josephson-CMOS systems. in *IEEE Transactions on Applied Superconductivity* (2005). doi:10.1109/TASC.2005.849786.
27. Henkels, W. H. *et al.* A low temperature 12 ns DRAM. in *International Symposium on VLSI Technology, Systems and Applications* 32–35 (IEEE). doi:10.1109/VTSA.1989.68576.
28. Henkels, W. H. *et al.* Low temperature SER and noise in a high speed DRAM. in *Proceedings of the Workshop on Low Temperature Semiconductor Electronics* (1989). doi:10.1109/ltse.1989.50171.
29. Mohler, R. L. *et al.* A 4-Mb Low-Temperature DRAM. *IEEE J. Solid-State Circuits* **26**, 1519–1529 (1991).
30. Vogelsang, T. Understanding the energy consumption of Dynamic Random Access Memories. in *Proceedings of the Annual International Symposium on Microarchitecture, MICRO* 363–374 (2010). doi:10.1109/MICRO.2010.42.
31. Mitchell, C., McCartney, C. L., Hunt, M. & Ho, F. D. Characteristics of a three-transistor DRAM circuit utilizing a

ferroelectric transistor. in *Integrated Ferroelectrics* vol. 157 31–38 (Taylor and Francis Ltd., 2014).
32. Tack, M. R., Gao, M., Claeys, C. L. & Declerck, G. J. The Multistable Charge-Controlled Memory Effect in SOI MOS Transistors at Low Temperatures. *IEEE Trans. Electron Devices* (1990) doi:10.1109/16.108200.
33. Morishita, F. *et al.* Leakage mechanism due to floating body and countermeasure on dynamic retention mode of SOI-DRAM. in *Digest of Technical Papers - Symposium on VLSI Technology* (1995). doi:10.1109/vlsit.1995.520897.
34. Ohsawa, T. *et al.* A Memory Using One-transistor Gain Cell on SOI(FBC) with Performance Suitable for Embedded DRAM's. in *IEEE Symposium on VLSI Circuits, Digest of Technical Papers* (2003). doi:10.1109/vlsic.2003.1221171.
35. Collaert, N. *et al.* A low-voltage biasing scheme for aggressively scaled bulk FinFET 1T-DRAM featuring 10s retention at 85°C. in *Digest of Technical Papers - Symposium on VLSI Technology* (2010). doi:10.1109/VLSIT.2010.5556211.
36. Park, K. H., Park, C. M., Kong, S. H. & Lee, J. H. Novel double-gate 1T-DRAM cell using nonvolatile memory functionality for high-performance and highly scalable embedded DRAMs. *IEEE Trans. Electron Devices* (2010) doi:10.1109/TED.2009.2038650.
37. Bae, J. H. *et al.* Characterization of a Capacitorless DRAM Cell for Cryogenic Memory Applications. *IEEE Electron Device Lett.* (2019) doi:10.1109/LED.2019.2933504.
38. Song, Y. J., Jeong, G., Baek, I. G. & Choi, J. What lies ahead for resistance-based memory technologies? *Computer (Long. Beach. Calif).* **46**, 30–36 (2013).
39. Lee, H. Y. *et al.* Evidence and solution of over-RESET problem for HfOX based resistive memory with sub-ns switching speed and high endurance. in *Technical Digest - International Electron Devices Meeting, IEDM* (2010). doi:10.1109/IEDM.2010.5703395.
40. Govoreanu, B. *et al.* 10×10nm 2 Hf/HfO x crossbar resistive RAM with excellent performance, reliability and low-energy operation. in *Technical Digest - International Electron Devices Meeting, IEDM* (2011). doi:10.1109/IEDM.2011.6131652.
41. Ielmini, D. *et al.* Scaling analysis of submicrometer nickel-oxide-based resistive switching memory devices. *J. Appl. Phys.* (2011) doi:10.1063/1.3544499.
42. Kügeler, C., Zhang, J., Hoffmann-Eifert, S., Kim, S. K. & Waser, R. Nanostructured resistive memory cells based on 8-nm-thin TiO2 films deposited by atomic layer deposition. *J. Vac. Sci. Technol. B, Nanotechnol. Microelectron. Mater. Process. Meas. Phenom.* (2011) doi:10.1116/1.3536487.
43. Walczyk, C. *et al.* On the role of Ti adlayers for resistive switching in HfO2-based metal-insulator-metal structures: Top versus bottom electrode integration. *J. Vac. Sci. Technol. B, Nanotechnol. Microelectron. Mater. Process. Meas. Phenom.* (2011) doi:10.1116/1.3536524.
44. Hirose, S., Nakayama, A., Niimi, H., Kageyama, K. & Takagi, H. Resistance switching and retention behaviors in polycrystalline La-doped SrTiO3 ceramics chip devices. *J. Appl. Phys.* (2008) doi:10.1063/1.2975316.
45. Yao, J., Sun, Z., Zhong, L., Natelson, D. & Tour, J. M. Resistive switches and memories from silicon oxide. *Nano Lett.* (2010) doi:10.1021/nl102255r.
46. Lörtscher, E., Ciszek, J. W., Tour, J. & Riel, H. Reversible and controllable switching of a single-molecule junction. *Small* (2006) doi:10.1002/smll.200600101.
47. Walczyk, C. *et al.* Impact of temperature on the resistive switching behavior of embedded HfO2-based RRAM devices. *IEEE Trans. Electron Devices* (2011) doi:10.1109/TED.2011.2160265.
48. Ahn, C. *et al.* Temperature-dependent studies of the electrical properties and the conduction mechanism of HfOx-based RRAM. in *Proceedings of Technical Program - 2014 International Symposium on VLSI Technology, Systems and Application, VLSI-TSA 2014* (2014). doi:10.1109/VLSI-TSA.2014.6839685.
49. Shang, J. *et al.* Thermally stable transparent resistive random access memory based on all-oxide heterostructures. *Adv. Funct. Mater.* (2014) doi:10.1002/adfm.201303274.
50. Fang, R., Chen, W., Gao, L., Yu, W. & Yu, S. Low-temperature characteristics of HfO. *IEEE Electron Device Lett.* **36**, 567–569 (2015).
51. Blonkowski, S. & Cabout, T. Bipolar resistive switching from liquid helium to room temperature. *J. Phys. D. Appl. Phys.* (2015) doi:10.1088/0022-3727/48/34/345101.
52. Takashima, D. Overview of FeRAMs: Trends and perspectives. *2011 11th Annu. Non-Volatile Mem. Technol. Symp. NVMTS 2011* 36–41 (2011) doi:10.1109/NVMTS.2011.6137107.
53. Trentzsch, M. *et al.* A 28nm HKMG super low power embedded NVM technology based on ferroelectric FETs. in *Technical Digest - International Electron Devices Meeting, IEDM* (2017). doi:10.1109/IEDM.2016.7838397.
54. Dünkel, S. *et al.* A FeFET based super-low-power ultra-fast embedded NVM technology for 22nm FDSOI and beyond. in *Technical Digest - International Electron Devices Meeting, IEDM* (2018). doi:10.1109/IEDM.2017.8268425.
55. Chatterjee, K. *et al.* Self-Aligned, Gate Last, FDSOI, Ferroelectric Gate Memory Device with 5.5-nm Hf0.8Zr0.2O2, High Endurance and Breakdown Recovery. *IEEE Electron Device Lett.* (2017) doi:10.1109/LED.2017.2748992.
56. Florent, K. *et al.* Vertical Ferroelectric HfO 2 FET based on 3-D NAND Architecture: Towards Dense Low-Power Memory. in *Technical Digest - International Electron Devices Meeting, IEDM* (2019). doi:10.1109/IEDM.2018.8614710.
57. Rowley, S. E. *et al.* Ferroelectric quantum criticality. *Nat. Phys.* (2014) doi:10.1038/nphys2924.
58. Meng, X. J. *et al.* Temperature dependence of ferroelectric and dielectric properties of PbZr0.5Ti0.5O3 thin film based

capacitors. *Appl. Phys. Lett.* **81**, 4035–4037 (2002).
59. Wang, Z. et al. Cryogenic Characterization of Antiferroelectric Zirconia down to 50 mK. in *Device Research Conference - Conference Digest, DRC* (2019). doi:10.1109/DRC46940.2019.9046475.
60. Wang, Z. et al. Cryogenic characterization of a ferroelectric field-effect-transistor. *Appl. Phys. Lett.* (2020) doi:10.1063/1.5129692.
61. Na, T., Kang, S. H. & Jung, S. O. STT-MRAM Sensing: A Review. *IEEE Trans. Circuits Syst. II Express Briefs* **68**, 12–18 (2021).
62. Lang, L. et al. A low temperature functioning CoFeB/MgO-based perpendicular magnetic tunnel junction for cryogenic nonvolatile random access memory. *Appl. Phys. Lett.* **116**, (2020).
63. Rowlands, G. E. et al. A cryogenic spin-torque memory element with precessional magnetization dynamics. *Sci. Rep.* (2019) doi:10.1038/s41598-018-37204-3.
64. Yau, J. B., Fung, Y. K. K. & Gibson, G. W. Hybrid cryogenic memory cells for superconducting computing applications. in *2017 IEEE International Conference on Rebooting Computing, ICRC 2017 - Proceedings* (2017). doi:10.1109/ICRC.2017.8123684.
65. Alam, S., Hossain, M. S. & Aziz, A. A non-volatile cryogenic random-access memory based on the quantum anomalous Hall effect. *Sci. Rep.* **11**, 1–9 (2021).
66. Serlin, M. et al. Intrinsic quantized anomalous Hall effect in a moiré heterostructure. *Science (80-. ).* **367**, 900–903 (2020).
67. Ovshinsky, S. R. Reversible electrical switching phenomena in disordered structures. *Phys. Rev. Lett.* **21**, 1450–1453 (1968).
68. Lai, S. Current status of the phase change memory and its future. in *Technical Digest - International Electron Devices Meeting* 255–258 (2003). doi:10.1109/iedm.2003.1269271.
69. Yi, H. T., Choi, T. & Cheong, S. W. Reversible colossal resistance switching in (La,Pr,Ca) MnO3: Cryogenic nonvolatile memories. *Appl. Phys. Lett.* (2009) doi:10.1063/1.3204690.
70. Tahara, S., Ishida, I., Ajisawa, Y. & Wada, Y. Experimental vortex transitional nondestructive read-out Josephson memory cell. *J. Appl. Phys.* (1989) doi:10.1063/1.343077.
71. Tahara, S. et al. 4-Kbit Josephson Nondestructive ReadOut Ram Operated At 580 psec and 6.7 mW. *IEEE Trans. Magn.* (1991) doi:10.1109/20.133751.
72. Alam, S., Jahangir, M. A. & Aziz, A. A Compact Model for Superconductor- Insulator-Superconductor (SIS) Josephson Junctions. *IEEE Electron Device Lett.* **41**, 1249–1252 (2020).
73. Nagasawa, S., Numata, H., Hashimoto, Y. & Tahara, S. High-frequency clock operation of josephson 256-word x 16-bit rams. *IEEE Trans. Appl. Supercond.* (1999) doi:10.1109/77.783834.
74. Nagasawa, S., Hinode, K., Satoh, T., Kitagawa, Y. & Hidaka, M. Design of all-dc-powered high-speed single flux quantum random access memory based on a pipeline structure for memory cell arrays. *Supercond. Sci. Technol.* (2006) doi:10.1088/0953-2048/19/5/S34.
75. Nagasawa, S., Hashimoto, Y., Numata, H. & Tahara, S. A 380 ps, 9.5 mW Josephson 4-Kbit RAM Operated at a High Bit Yield. *IEEE Trans. Appl. Supercond.* (1995) doi:10.1109/77.403086.
76. Kirichenko, A., Mukhanov, O., Kirichenko, A. F., Mukhanov, O. A. & Brock, D. K. A Single Flux Quantum Cryogenic Random Access Memory Rapid Single Flux Quantum Digital Electronics View project Inductance extraction and flux trapping analysis of superconducting circuits View project A Single Flux Quantum Cryogenic Random Access Memory. in *Extended Abstracts of 7th International Superconducting Electronics Conference (ISEC'99)* 124–127 (Proc. Ext. Abstracts 7th Int. Supercond. Electron. Conf. (ISEC'99), 1999).
77. Kirichenko, A. F., Sarwana, S., Brock, D. K. & Radpavar, M. Pipelined DC-powered SFQ RAM. in *IEEE Transactions on Applied Superconductivity* (2001). doi:10.1109/77.919401.
78. Yuh, P. F. A Buffered Nondestructive-Readout Josephson Memory Cell with Three Gates. *IEEE Trans. Magn.* (1991) doi:10.1109/20.133809.
79. Yuh, P. F. A 2-kbit Superconducting Memory Chip. *IEEE Trans. Appl. Supercond.* **3**, 3013–3021 (1993).
80. Polonsky, S. V., Kirichenko, A. F., Semenov, V. K. & Likharev, K. K. Rapid Single Flux Quantum Random Access Memory. *IEEE Trans. Appl. Supercond.* (1995) doi:10.1109/77.403223.
81. Polyakov, Y., Narayana, S. & Semenov, V. K. Flux trapping in superconducting circuits. *IEEE Trans. Appl. Supercond.* **17**, 520–525 (2007).
82. Narayana, S., Polyakov, Y. A. & Semenov, V. K. Evaluation of flux trapping in superconducting circuits. *IEEE Trans. Appl. Supercond.* **19**, 640–643 (2009).
83. Jackman, K. & Fourie, C. J. Flux Trapping Analysis in Superconducting Circuits. *IEEE Trans. Appl. Supercond.* **27**, (2017).
84. Holmes, D. S., Ripple, A. L. & Manheimer, M. A. Energy-Efficient Superconducting Computing—Power Budgets and Requirements. *IEEE Trans. Appl. Supercond.* **23**, 1701610–1701610 (2013).
85. Manheimer, M. A. Cryogenic computing complexity program: Phase 1 introduction. *IEEE Trans. Appl. Supercond.* **25**, (2015).
86. Tolpygo, S. K. Superconductor digital electronics: Scalability and energy efficiency issues. *Low Temperature Physics*

(2016) doi:10.1063/1.4948618.
87. Tolpygo, S. K. *et al.* Inductance of circuit structures for MIT LL superconductor electronics fabrication process with 8 niobium layers. *IEEE Trans. Appl. Supercond.* **25**, (2015).
88. Braiman, Y., Nair, N., Rezac, J. & Imam, N. Memory cell operation based on small Josephson junctions arrays. *Supercond. Sci. Technol.* **29**, 124003 (2016).
89. Braiman, Y., Neschke, B., Nair, N., Imam, N. & Glowinski, R. Memory states in small arrays of Josephson junctions. *Phys. Rev. E* (2016) doi:10.1103/PhysRevE.94.052223.
90. Nair, N., Jafari-Salim, A., D'Addario, A., Imam, N. & Braiman, Y. Experimental demonstration of a Josephson cryogenic memory cell based on coupled Josephson junction arrays. *Supercond. Sci. Technol.* **32**, 115012 (2019).
91. Nair, N. & Braiman, Y. A ternary memory cell using small Josephson junction arrays. *Supercond. Sci. Technol.* (2018) doi:10.1088/1361-6668/aae2a9.
92. Miloshevsky, A., Nair, N., Imam, N. & Braiman, Y. High-Tc Superconducting Memory Cell. *J. Supercond. Nov. Magn.* 1–10 (2021) doi:10.1007/S10948-021-06069-5/FIGURES/8.
93. Hilgenkamp, H. Josephson Memories. *J. Supercond. Nov. Magn.* **34**, 1621–1625 (2020).
94. Wada, Y. Josephson Memory Technology. *Proc. IEEE* **77**, 1194–1207 (1989).
95. Hidaka, Y. Superconductor magnetic memory using magnetic films. 656 (1991).
96. Ryazanov, V. V. *et al.* Magnetic josephson junction technology for digital and memory applications. in *Physics Procedia* (2012). doi:10.1016/j.phpro.2012.06.126.
97. Oboznov, V. A., Bol'ginov, V. V., Feofanov, A. K., Ryazanov, V. V. & Buzdin, A. I. Thickness dependence of the Josephson ground states of superconductor- ferromagnet-superconductor junctions. *Phys. Rev. Lett.* (2006) doi:10.1103/PhysRevLett.96.197003.
98. Shelukhin, V. *et al.* Observation of periodic π-phase shifts in ferromagnet-superconductor multilayers. *Phys. Rev. B - Condens. Matter Mater. Phys.* (2006) doi:10.1103/PhysRevB.73.174506.
99. Robinson, J. W. A., Piano, S., Burnell, G., Bell, C. & Blamire, M. G. Critical current oscillations in strong ferromagnetic π junctions. *Phys. Rev. Lett.* (2006) doi:10.1103/PhysRevLett.97.177003.
100. Niedzielski, B. M., Gingrich, E. C., Loloee, R., Pratt, W. P. & Birge, N. O. S/F/S Josephson junctions with single-domain ferromagnets for memory applications. *Supercond. Sci. Technol.* (2015) doi:10.1088/0953-2048/28/8/085012.
101. Bol'ginov, V. V., Stolyarov, V. S., Sobanin, D. S., Karpovich, A. L. & Ryazanov, V. V. Magnetic switches based on Nb-PdFe-Nb Josephson junctions with a magnetically soft ferromagnetic interlayer. *JETP Lett. 2012 957* **95**, 366–371 (2012).
102. Glick, J. A., Loloee, R., Pratt, W. P. & Birge, N. O. Critical Current Oscillations of Josephson Junctions Containing PdFe Nanomagnets. *IEEE Trans. Appl. Supercond.* (2017) doi:10.1109/TASC.2016.2630024.
103. Dayton, I. M. *et al.* Experimental Demonstration of a Josephson Magnetic Memory Cell with a Programmable φ-Junction. *IEEE Magn. Lett.* **9**, (2018).
104. Ryazanov, V. V. *et al.* Coupling of two superconductors through a ferromagnet: Evidence for a π junction. *Phys. Rev. Lett.* (2001) doi:10.1103/PhysRevLett.86.2427.
105. Bannykh, A. A. *et al.* Josephson tunnel junctions with a strong ferromagnetic interlayer. *Phys. Rev. B - Condens. Matter Mater. Phys.* **79**, 054501 (2009).
106. Vernik, I. V. *et al.* Magnetic josephson junctions with superconducting interlayer for cryogenic memory. *IEEE Trans. Appl. Supercond.* (2013) doi:10.1109/TASC.2012.2233270.
107. Larkin, T. I. *et al.* Ferromagnetic Josephson switching device with high characteristic voltage. *Appl. Phys. Lett.* **100**, 222601 (2012).
108. Bakurskiy, S. V. *et al.* Theoretical model of superconducting spintronic SIsFS devices. *Appl. Phys. Lett.* **102**, 192603 (2013).
109. Bergeret, F. S., Volkov, A. F. & Efetov, K. B. Enhancement of the Josephson current by an exchange field in superconductor-ferromagnet structures. *Phys. Rev. Lett.* (2001) doi:10.1103/PhysRevLett.86.3140.
110. Krivoruchko, V. N. & Koshina, E. A. From inversion to enhancement of the dc Josephson current in (formula presented) tunnel structures. *Phys. Rev. B - Condens. Matter Mater. Phys.* (2001) doi:10.1103/PhysRevB.64.172511.
111. Golubov, A. A., Kupriyanov, M. Y. & Fominov, Y. V. Critical current in SFIFS junctions. *JETP Lett.* (2002) doi:10.1134/1.1475721.
112. Barash, Y. S., Bobkova, I. V. & Kopp, T. Josephson current in S-FIF-S junctions: Nonmonotonic dependence on misorientation angle. *Phys. Rev. B - Condens. Matter Mater. Phys.* (2002) doi:10.1103/PhysRevB.66.140503.
113. Bell, C. *et al.* Controllable Josephson current through a pseudospin-valve structure. *Appl. Phys. Lett.* **84**, 1153–1155 (2004).
114. Kontos, T. *et al.* Josephson Junction through a Thin Ferromagnetic Layer: Negative Coupling. *Phys. Rev. Lett.* (2002) doi:10.1103/PhysRevLett.89.137007.
115. Blum, Y., Tsukernik, A., Karpovski, M. & Palevski, A. Oscillations of the Superconducting Critical Current in Nb-Cu-Ni-Cu-Nb Junctions. *Phys. Rev. Lett.* (2002) doi:10.1103/PhysRevLett.89.187004.
116. Glick, J. A. *et al.* Critical current oscillations of elliptical Josephson junctions with single-domain ferromagnetic layers. *J. Appl. Phys.* (2017) doi:10.1063/1.4989392.

117. Baek, B., Rippard, W. H., Benz, S. P., Russek, S. E. & Dresselhaus, P. D. Hybrid superconducting-magnetic memory device using competing order parameters. *Nat. Commun.* (2014) doi:10.1038/ncomms4888.
118. Golod, T., Iovan, A. & Krasnov, V. M. Single Abrikosov vortices as quantized information bits. *Nat. Commun. 2015 61* **6**, 1–5 (2015).
119. Stoner, E. C. & Wohlfarth, E. P. A mechanism of magnetic hysteresis in heterogeneous alloys. *IEEE Trans. Magn.* (1991) doi:10.1109/TMAG.1991.1183750.
120. Goldobin, E. *et al.* Memory cell based on a φ Josephson junction. *Appl. Phys. Lett.* **102**, 242602 (2013).
121. McCaughan, A. N. & Berggren, K. K. A superconducting-nanowire three-terminal electrothermal device. *Nano Lett.* (2014) doi:10.1021/nl502629x.
122. Peotta, S. & Di Ventra, M. Superconducting Memristors. *Phys. Rev. Appl.* (2014) doi:10.1103/PhysRevApplied.2.034011.
123. Alam, S., Hossain, M. S. & Aziz, A. A cryogenic memory array based on superconducting memristors. *Appl. Phys. Lett.* **119**, 082602 (2021).
124. Stewart, W. C. Current-voltage characteristics of Josephson junctions. *Appl. Phys. Lett.* (1968) doi:10.1063/1.1651991.
125. McCumber, D. E. Effect of ac impedance on dc voltage-current characteristics of superconductor weak-link junctions. *J. Appl. Phys.* (1968) doi:10.1063/1.1656743.
126. Ingold, G. L., Grabert, H. & Eberhardt, U. Cooper-pair current through ultrasmall Josephson junctions. *Phys. Rev. B* (1994) doi:10.1103/PhysRevB.50.395.
127. Van Den Brink, A. M., Schön, G. & Geerligs, L. J. Combined single-electron and coherent-Cooper-pair tunneling in voltage-biased Josephson junctions. *Phys. Rev. Lett.* (1991) doi:10.1103/PhysRevLett.67.3030.
128. Suleiman, M., Sarott, M. F., Trassin, M., Badarne, M. & Ivry, Y. Nonvolatile voltage-tunable ferroelectric-superconducting quantum interference memory devices. *Appl. Phys. Lett.* **119**, 112601 (2021).
129. Ghoshal, U., Kroger, H. & Van Duzer, T. Superconductor-Semiconductor Memories. *IEEE Transactions on Applied Superconductivity* (1993) doi:10.1109/77.233542.
130. Feng, Y. J. *et al.* Josephson-CMOS hybrid memory with ultra-high-speed interface circuit. in *IEEE Transactions on Applied Superconductivity* (2003). doi:10.1109/TASC.2003.813902.
131. Duzer, T. Van, Liu, Q., Meng, X., Whiteley, S. & Yoshikawa, N. High-speed interface amplifiers for SFQ-to-CMOS signal conversion. in *International Superconductor Electronics Conference, ISEC* (2003).
132. Liu, Q. *et al.* Simulation and measurements on a 64-kbit hybrid josephson-CMOS memory. in *IEEE Transactions on Applied Superconductivity* (2005). doi:10.1109/TASC.2005.849863.
133. Liu, Q. *et al.* Latency and power measurements on a 64-kb hybrid Josephson-CMOS memory. in *IEEE Transactions on Applied Superconductivity* (2007). doi:10.1109/TASC.2007.898698.
134. Kuwabara, K., Jin, H., Yamanashi, Y. & Yoshikawa, N. Design and implementation of 64-kb CMOS static RAMs for Josephson-CMOS hybrid memories. *IEEE Trans. Appl. Supercond.* **23**, (2013).
135. Tanaka, M. *et al.* Josephson-CMOS Hybrid Memory with Nanocryotrons. *IEEE Trans. Appl. Supercond.* **27**, (2017).
136. Aziz, A. *et al.* Single-ended and differential MRAMs based on spin hall effect: A layout-aware design perspective. *Proc. IEEE Comput. Soc. Annu. Symp. VLSI, ISVLSI* **07-10-July-2015**, 333–338 (2015).
137. Nguyen, M. H. *et al.* Cryogenic Memory Architecture Integrating Spin Hall Effect based Magnetic Memory and Superconductive Cryotron Devices. *Sci. Rep.* (2020) doi:10.1038/s41598-019-57137-9.
138. Mukhanov, O. A., Kirichenko, A. F., Filippov, T. V. & Sarwana, S. Hybrid semiconductor-superconductor fast-readout memory for digital RF receivers. in *IEEE Transactions on Applied Superconductivity* (2011). doi:10.1109/TASC.2010.2089409.
139. Suzuki, H., Inoue, A., Imamura, T. & Hasuo, S. Josephson driver to interface Josephson junctions to semiconductor transistors. in *Technical Digest - International Electron Devices Meeting* 290–293 (Publ by IEEE, 1988). doi:10.1109/iedm.1988.32814.
140. Ghoshal, U., Kishore, S., Feldman, A., Huynh, L. & Van Duzer, T. CMOS amplifier designs for Josephson-CMOS interface circuits. *IEEE Trans. Appl. Supercond.* **5**, 2640–2643 (1995).